\documentclass[twocolumn,numbering,showpacs]{revtex4-1}

\usepackage{graphicx,color}
\usepackage{latexsym}
\usepackage{amsmath,amssymb}        
\usepackage[draft=false]{hyperref}
\usepackage{mathrsfs}
\usepackage{comment}
\usepackage{soul}
\usepackage{mathrsfs}
\definecolor{purple}{rgb}{0.58,0.0,0.83}
\definecolor{orange}{rgb}{1,0.5,0}
\DeclareSymbolFontAlphabet{\mathrsfs}{rsfs}
\DeclareMathAlphabet{\mathcal}{OMS}{cmsy}{m}{n}

\begin{document}

% -----> TITLE 

\title{Classification of a black hole spin out of its shadow using support vector machines}

% -----> AUTHORS 

\author{J. A. Gonz\'alez, F. S. Guzm\'an}
\affiliation{${}^{1}$ Laboratorio de Inteligencia Artificial y Superc\'omputo, Instituto de F\'{\i}sica y Matem\'aticas, Universidad Michoacana de San Nicol\'as de Hidalgo. Edificio C-3, Cd. Universitaria, 58040 Morelia, Michoac\'an, M\'exico.}
% --->   DATE

\date{\today}

% -----> ABSTRACT

\begin{abstract}
We use Support Vector Machines (SVMs) to classify the spin of a black hole. 
The SVMs are trained and tested with a catalog of numerically generated images of black holes, assuming disk and spherical matter models with monochromatic emission with wavelength of 4mm. 
We determine the accuracy  of the SVM to classify the spin in terms of the image resolution, for which we consider three resolutions of $16^2,~32^2$ and $64^2$ pixels. Our approach is applied to the specific mass of the Supermassive Black Hole (SMBH) at the center of the Milky Way. Our findings are that when the distribution is a thin disk, the accuracy in the classification is acceptable even for the coarsest resolution with accuracy over 90\%, whereas for the spherical distribution it drops below 80\% for low and intermediate resolutions. The results show how the distribution of matter, the angle of vision and the image resolution influence the difficulty to determine the correct range of black hole spin.
\end{abstract}

% ----->   PACS

% ----->   MAKETITLE 

\maketitle

% ----->     INTRODUCTION  
\section{Introduction}
\label{sec:introduction}

It is exciting that very soon the Event Horizon Telescope (EHT) will provide the first images ever, of horizon size scale observations of the two most studied supermassive black holes Sgr A$^{*}$ \cite{Doeleman2008,Doeleman2011,Doeleman2012} and that in M87 \cite{Chatzopoulos}. The accessible information obtained from the observations will reveal the most interesting features of the physics developing near the black hole horizon, including tests of conventional models of matter revolving around the black hole \cite{Silk}, the magnetic fields involved \cite{Dexteretal2010}, radiation-matter coupling models happening at very strong gravitational fields \cite{matterradiation}, quantum effects like Hawking radiation processes \cite{Hawking1}, tests of General Relativity in such strong field scenario \cite{PsaltisReview}, among other applications.

The fingerprint of a black hole is expected to be reconstructed out of the image  produced by the matter around it.  This situation defines an inverse problem for both, the parameters of the black hole and the matter whose image is expected to be seen. This problem consists in the reconstruction of  the intrinsic parameters, namely spin and mass of the black hole, spatial distribution and nature of the matter producing the radiation that will be captured by the EHT, the magnetic field configuration and a radiation-matter interaction model, altogether in a possibly dynamical scenario \cite{Psaltis2018}. Also the inverse problem would  involve extrinsic parameters related to the orientation of the hole and the matter distribution with respect to the observer.

Extensive analyses have been performed that involve the General Relativistic Magnetohydrodynamics simulation of plasmas around the black hole in order to explore the effects of spin on the distribution and therefore images of the matter \cite{Dexteretal2010} including highly asymmetric configurations like tilted disks \cite{DexterFragile2013}, that solve the direct problem under a variety of conditions, in order to have a catalog of possible scenarios that could match the images obtained with the EHT, and  help constraining the matter models \cite{Psaltis2015}. In more dynamical scenarios compared with Sgr A${}^{*}$, similar analyses based on General Relativistic Hydrodynamics simulations have been used to determine, with a given uncertainty, the velocity of wandering black holes moving at supersonic speed, and the adiabatic index of the gas around it \cite{GonzalezGuzman2018} that could be applied to off-set SMBHs like the radio-loud QSO 3C 186 \cite{Chiaberge}.

Among the various parameters of  a SMBH system that need to be inferred from images, we investigate whether the spin of the black hole can be identified provided a very poor resolution image using Support Vector Machine  methods. As a first approach we consider a very simple situation, in which we assume the matter is distributed either as a disk or has a spherical profile. We also assume the radiation model corresponds to black body radiation but study the images in the single 4mm wavelength. 
Another restriction is that the matter and the black hole are assumed to be stationary, which is a zeroth order approximation of scenarios with variability \cite{Psaltis2018}.
Even with these many restrictions, determining the spin is expected to be challenging since it is known the images are not considerably influenced by the spin \cite{JohannsenPsaltis2010}, thus having an efficient method that can allow an estimate the spin could be useful. 

About the method, SVM are artificial intelligence machine learning methods used to classify parameter values, spin in our case, involving at least two phases, training and prediction. In this paper the training and prediction sets are  constructed using mock images of matter around black holes that we use to train and test the classification method. These images correspond to matter around the black hole intentionally using a rather poor resolution, expecting this to be the case of SMBHs farther than the M87, that cover a zone of a few horizon radii as expected for the EHT observations \cite{Doeleman2008,Doeleman2012}. In the end we want to investigate how the prediction capacity of SVMs is restricted either by image resolution or matter model.

The paper is written with the following structure:
In Sec~\ref{sec:two} we describe the construction of the catalog of images around the black hole.
In Sec~\ref{sec:three} we describe the methods and parameters of the SVM used to classify the images.
In Sec~\ref{sec:four} we show the results in terms of classification accuracy. Finally in section \ref{sec:five} we draw some final conclusions.

% ----->     IMAGES
\section{Images}
\label{sec:two}

The spin of the black hole is classified based on a catalog of images produced for simple radiation models and two  distributions of matter, namely a disk-like distribution and a spherical distribution. The construction of images assumes the matter radiates as a black body, and we consider images corresponding to a single frequency near the upper wavelength limit of the LMT band \cite{LMT}, specifically $\nu=7.495\times 10^{10}$Hz, that is, a wavelength of $\sim$4mm.

We assume the mass of the black hole to be $4.31\times 10^6 M_{\odot}$, the one estimated for the SMBH at Sgr A${}^{*}$ and the mock images are constructed using efficient ray tracing methods, for which we  specifically  use  GRTrans \cite{GRTrans}, which is much more efficient than our own tracer used in \cite{GonzalezGuzman2018}.

For the generation of the catalog we assume the orientation of the observer with respect to the equatorial plane of the black hole is unknown, and we consider that for the same value of the spin, a set of possible different images can be constructed. The orientation is given by the cosine of the angle $\theta$ formed by the observer's line of sight and a normal vector to the equatorial plane of the black hole, thus the parameter is $\mu=\cos(\theta)$.

We vary this parameter in the range between a nearly edge view and top head angle view $\mu\in[0.1,1.0]$.
The spin parameter $s$ of the black hole is assumed to lie within the interval $s\in[-0.99,+0.99]$. In this way, the catalog contains a wide range of values containing prograd and retrograd disk rotation, with values near the extreme possible values of spin.

\begin{figure}
\centering
\includegraphics[width= 4cm]{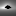}
\includegraphics[width= 4cm]{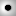}
\includegraphics[width= 4cm]{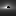}
\includegraphics[width= 4cm]{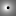}
\caption{In the top we show the images for spin parameter $s=-0.75$ and  angles of vision $\mu=0.21$ (left) and $\mu = 0.89$ (right). In the bottom the images presented have the same orientations but spin parameter $s=0.75$. These four images use resolution of 16$\times$16 pixels.}
\label{fig:one}
\end{figure}

\begin{figure}
\centering
\includegraphics[width= 4cm]{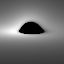}
\includegraphics[width= 4cm]{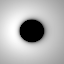}
\includegraphics[width= 4cm]{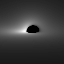}
\includegraphics[width= 4cm]{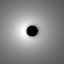}
\caption{Images of the system with the same parameters as in the previous figure but in this case using image resolution of 64$\times$64 pixels.}
\label{fig:two}
\end{figure}

In Figure \ref{fig:one} we show the images of the disk distribution with resolution 16$\times 16$ pixels, for two values of the spin and two different orientations. This is the type of images the SVM method should be able to classify, even though the resolution is very poor.
For comparison of the possible image resolution, we show in Figure \ref{fig:two} the same configurations but this time using resolution 64$\times$64 pixels. Figures \ref{fig:one} and \ref{fig:two} show the contrast that can be found in terms of resolution of images. As a byproduct we may learn how resolution and matter model affects the appropriate classification of the spin by the SVM.

% ----->     CLASSIFICATION
\section{Classification method}
\label{sec:three}

The SVM method consists in determining the optimal decision function solving the optimization problem \cite{CC01a,Abe}

\begin{equation}
\min\limits_{\vec{w},b,\vec{\xi}} \left(  \frac{1}{2}\vec{w}\cdot\vec{w} + C \sum_{i=1}^{l} \xi_i \right) ~,
\label{eq:svm_min}
\end{equation}

\noindent subject to the constraint

\begin{equation}
y_i\left[\vec{w}\cdot\vec{\phi}(\vec{x}_i)+b\right]\geq 1-\xi_i \,\,\,,
\label{eq:svm_cons}
\end{equation}

\noindent where $i=1,...,l$ selects one of the $l$ training input vectors, each one of them an $n-$dimensional vector $\vec{x}_i$ belonging to two possible classes $y_i=\{ 1,-1\}$. The parameters of the decision function to be optimized are $\vec{w}\in\Re^n$, the bias $b\in\Re$ and the non-negative slack variables to allow separability $\xi_i\geq0$, parameters known as $L_1$ soft margin SVMs. Finally, the free parameters are the mapping function $\vec{\phi}:\Re^n\rightarrow\Re^l $ which maps the $n-$dimensional inputs $\vec{x}_i$ into a $l-$dimensional feature subspace and $C$ is the penalty parameter of the error term. 

Instead of solving the problem defined by equations (\ref{eq:svm_min}-\ref{eq:svm_cons}) 
it is converted into an unconstrained problem using non-negative Lagrange multipliers and the optimal solution is found by solving the Karush-Kuhn-Tucker conditions \cite{Abe}. 
We solve the dual problem in the feature subspace with the kernel function associated with the mapping function via $K(\vec{x},\vec{y})=\vec{\phi}(\vec{x})\cdot\vec{\phi}(\vec{y})$.  

The particular kernel used in our study is the radial basis function kernel defined by

\begin{equation}
K(\vec{x}_i,\vec{x}_j)=e^{-\gamma || \vec{x}_i - \vec{x}_j ||^2}\,\,\,,
\end{equation}

\noindent with $\gamma\geq0$. Considering our problem is a multi-class one (lets say $m-$class) instead of a two-class problem, we use the {\it one-against all} approach classifying into $m(m-1)/2$ two-class problems and using a voting strategy. In case two classes have identical votes, the class appearing first in the array is chosen. For the analysis in this paper we use the library libSVM \cite{CC01a}.

% ----->     RESULTS
\section{Results}
\label{sec:four}

As described above, the spin values  we explore are within the interval $s\in[s_{min},s_{max}]$ with $s_{min}=-0.99$, $s_{max}=0.99$, and different orientations in the range $\mu\in[\mu_{min},\mu_{max}]$, where $\mu_{min}=0.1$ and $\mu_{max}=1$. Before explaining the classification process, let us describe the training set used in our analysis.

{\it Training set.} We use  50 values of the spin parameter $s$  defined by $s_i=s_{min}+i\Delta s$ for $i=0,...,49$ with $\Delta s=(s_{max}-s_{min})/49$. For each of these values we prepare 50 images corresponding to 50 different values of $\mu$ given by $\mu_j=\mu_{min}+j\Delta \mu$ with $\Delta \mu = (\mu_{max}-\mu_{min})/49$ and $j=0,...,49$. This means that the training set consists of a total of 2500 images for each of the two matter models.

{\it Classes.} In order to associate a spin with the black hole appearing in each image, we generate ten possible intervals for the spin and relate each one of them with one of the ten classes that the SVM uses to address the classification. 
The ten classes are defined by $C_1=[s_{min},s_{min}+(s_{max}-s_{min})/10)$, class $C_2=[s_{min}+(s_{max}-s_{min})/10,s_{min}+2(s_{max}-s_{min})/10$,...,$C_{10}=[s_{min}+9(s_{max}-s_{min})/10,s_{max}]$. 

The training set is used to obtain the optimal parameters of the SVM in order to classify the images of the set into one of the ten  classes $\{C_1,...,C_{10}\}$ associated with the spin range of the black hole. Once it is trained, the machine is used to classify the prediction set. In Table \ref{table:disk} we present the accuracy of the SVM classification for the disk-like and spherical distributions using the training set. The first column contains the three image resolutions used in our analysis, the second and third columns show the percentage of correctly classified images for the disk and spherical training sets.

\begin{table}[htbp]
\begin{center}
\begin{tabular}{|c|c|c|}
\hline
Resolution & Disk distribution & Spherical distribution \\
\hline
16x16 & 97.92 & 65.40 \\
\hline
32x32 & 97.92 & 85.08 \\
\hline
64x64 & 98.00 & 93.84 \\
\hline
\end{tabular}
\caption{Percentage of images of the training set correctly classified for the disk-like and spherical distributions  as function of image resolution.}
\label{table:disk}
\end{center}
\end{table}

%\textcolor{red}{Once we change the matter distribution, the results change considerably. Using a spherical distribution of matter provided by GRTrans \cite{GRTrans}, we can observe a big difference in the performance of the SVM depending on the image resolution. In Table \ref{table:sph} we present the corresponding percentages of correctly classified images for the spherical distribution and we can observe how big  the difference is between the low and highest resolution. }

{\it Global prediction set.} This set is prepared with 15 values of $s$ and 15 values of $\mu$ uniformly distributed in the intervals $\hat{s}\in[\hat{s}_{min},\hat{s}_{max}]$ and $\hat{\mu}\in [\mu_{min},\mu_{max}]$. In total this set contains 225 images.
Notice the interval of the orientation is the same as for the training set whereas the interval for the spin is different since we want to avoid the values of the spin that repeat with the values in the training set. The specific values of the spin are $\hat{s}_a=\hat{s}_{min}+a\Delta\hat{s}$ where $\Delta \hat{s}=(\hat{s}_{min}-\hat{s}_{max})/14$ for $a=0,...,14$ where $\hat{s}_{min}=-\hat{s}_{max}=-0.9702$. The values for the orientation are $\hat{\mu}=\mu_{min}+b\Delta \hat{\mu}$ where $\Delta \hat{\mu}=(\mu_{max}-\mu_{min})/14$ and $b=0,...,14$, where the minimum and maximum of $\mu$ are the same as for the training set.  The results are presented in Table \ref{tab:tab2}. For the disk the SVM  is insensitive to the change of resolution obtaining accuracies over 96\% in all the cases, whereas for the spherical distribution the accuracy is not as good as for disks, but grows with image resolution.

These results are obtained using a prediction set containing values within the whole spin and orientation domains. We would like to know how the classification accuracy depends on the orientation parameter. For this it is necessary to define prediction sets for restricted values of $\mu$.

\begin{table}[htbp]
\begin{center}
\begin{tabular}{|c|c|c|}
\hline
Resolution & Disk distribution & Spherical distribution \\
\hline
16x16 & 97.33 & 57.33 \\
\hline
32x32 & 97.78 & 67.56 \\
\hline
64x64 & 96.44 & 82.67 \\
\hline
\end{tabular}
\caption{Percentage of images of the global prediction set correctly classified for the disk-like and spherical distributions for as function of image resolution.}
\label{tab:tab2}
\end{center}
\end{table}

{\it $\mu-$dependent prediction sets.} These sets are prepared with $N$ different values of $s$ uniformly distributed in the whole spin domain $\hat{s}\in[\hat{s}_{min},\hat{s}_{max}]$. The specific values of spin in this set are $\hat{s}_a=\hat{s}_{min}+a\Delta\hat{s}$ where $\Delta \hat{s}=(\hat{s}_{min}-\hat{s}_{max})/(N-1)$ $a=0,...,N-1$, where $\hat{s}_{min}=-\hat{s}_{max}=-0.985$. In order to determine the dependency of the prediction as function of $\mu$, we divide the domain of $[\mu_{min},\mu_{max}]$ in the following nine subsets $\mu_i = [0.1\times i,0.1\times (i+1)]$ for $i=1,...,9$. In each of these subsets we pick 30 uniformly distributed values of $\mu$. Then for each subset $\mu_i$ the number of images in the prediction set is $30\times N$.

In a first experiment we used $N=30$. Then for each subdomain $\mu_i$ the prediction set consists in 900 images. In this particular exercise, the spin values are sufficiently separated from each other, and from the boundaries between consecutive classes of spin, that accuracy in prediction is 100\% for all the image resolutions and for the disk and spherical distributions of matter. In order for the set to contain sufficient cases near the boundaries between consecutive spin classes we use $N=120$, which makes $\Delta \hat{s}=0.01642$, which is smaller than the separation between spin values in the training set and therefore the classification can fail in various cases. These are the training sets used in the analysis below.

\begin{figure}
\centering
\includegraphics[width= 8cm]{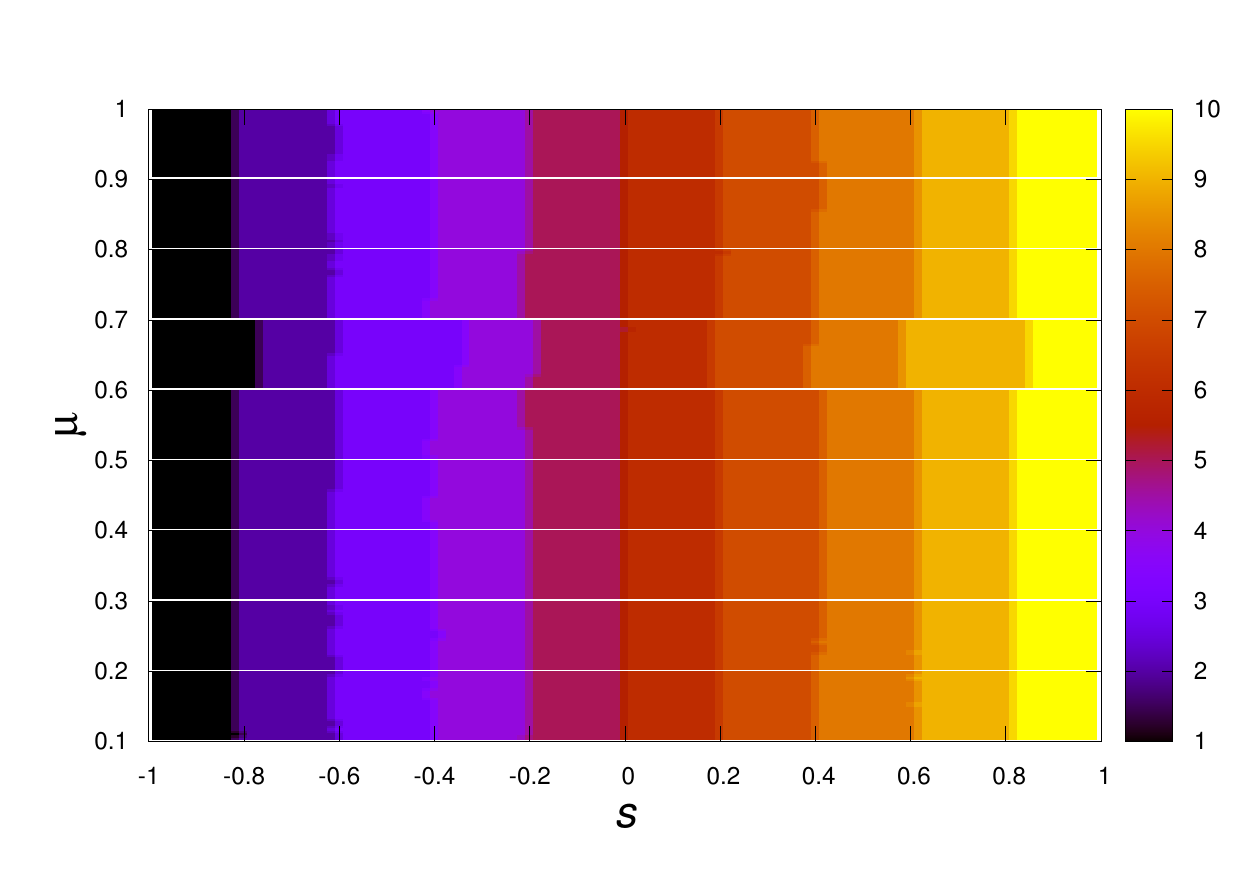}
\caption{Classification of the prediction set for the various subdomains of the inclination parameter $\mu$ and the disk distribution case. For this plot we use the images with resolution of $32\times 32$ pixels.}
\label{fig:Dependence_mu}
\end{figure}

In Figure \ref{fig:Dependence_mu} we represent the accuracy on the prediction sets for the images generated by disk distributions using the intermediate image resolution $32\times 32$ pixels. Keeping in mind that the classification refers to the value of the spin, the horizontal axis of this Figure corresponds to the spin value. Each of the 10 classes $C_i$ of spin has a color associated. For example, if an image in the prediction is classified within class $C_1$ by the SVM we use black, whereas if the images is classified within class $C_{10}$ we use yellow. We present these results for each of the 9 subdomains of orientation $\mu_1,...,\mu_9$ in the same plot.
The first interesting result is that the SVM has  classification errors between neighboring classes. For a particular case, various images corresponding to class $C_3$ are classified in class $C_2$, independently of the range of $\mu$. Something similar happens between classes $C_7$ and $C_8$, and so on. Nevertheless the bands appear well defined between classes, which indicates that the prediction fails only near the boundaries between classes. The second interesting result is that for the specific orientation domain $\mu_6 = [0.6,0.7]$ the error in the classification is clearly bigger than for the other ranges. This implies that for this intermediate range of vision angle, the classification is particularly innaccurate.

\begin{figure}
\centering
\includegraphics[width= 8cm]{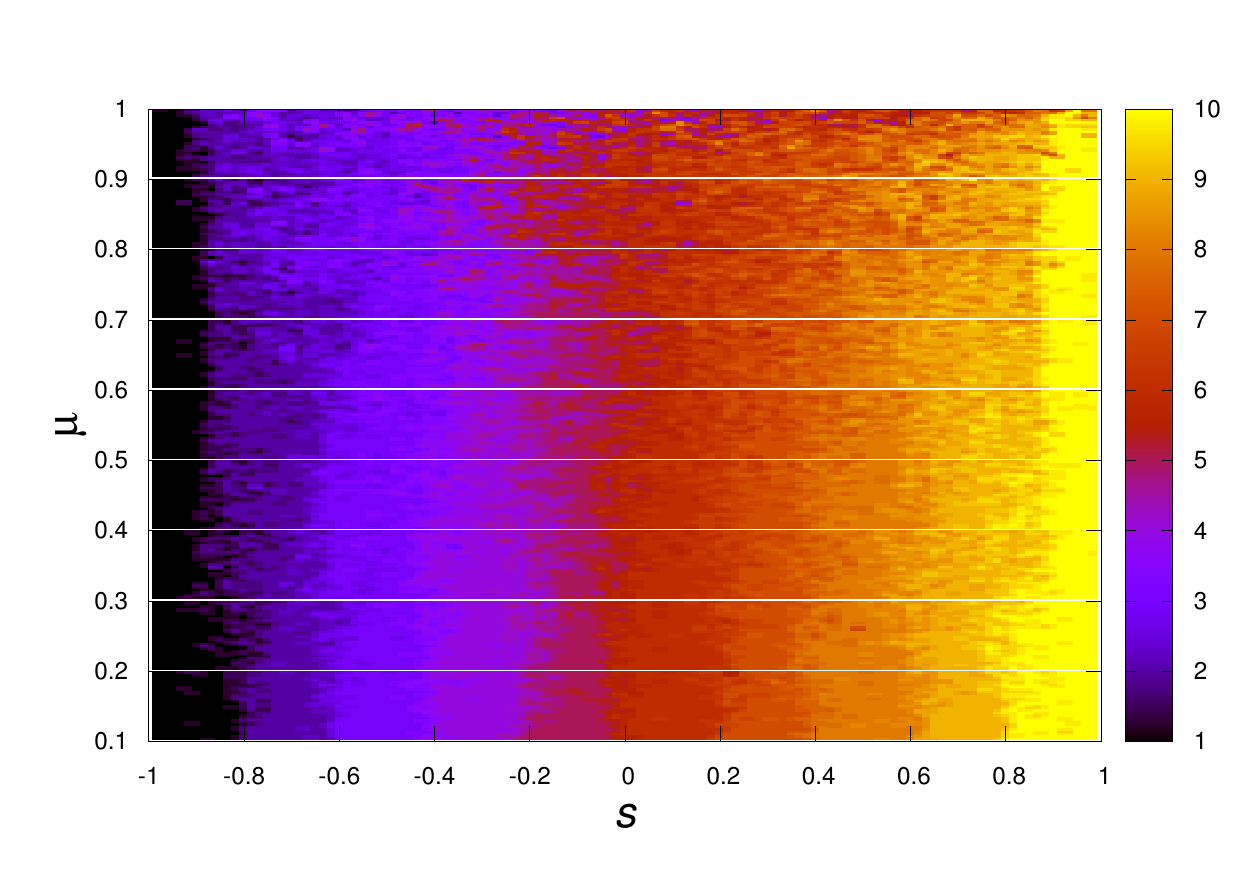}
\caption{Classification of the prediction set for the various subdomains of the inclination parameter $\mu$ and the spherical distribution case. For this plot we use the images with resolution of $32\times 32$ pixels.}
\label{fig:Dependence_spin_sph}
\end{figure}

In Figure \ref{fig:Dependence_spin_sph} we present  the results of classification for the prediction set generated by spherical distributions of matter using the same code of colors as for the disks. In this case the SVM has errors of correct classification between all the boundaries between each pair of consecutive classes. The first orientation domain $\mu_1=[0.1,0.2]$ is the case with less errors. This Figure illustrates how the prediction looses quality for bigger values of $\mu$. It is worth noticing that the color bands are not as clearly defined as for the disk distribution. We consider that one possible reason for this lack of precision is that the spherical symmetry of the matter distribution increases the degeneracy of images, making the appropriate classification a lot harder.

\begin{figure}
\centering
\includegraphics[width= 8cm]{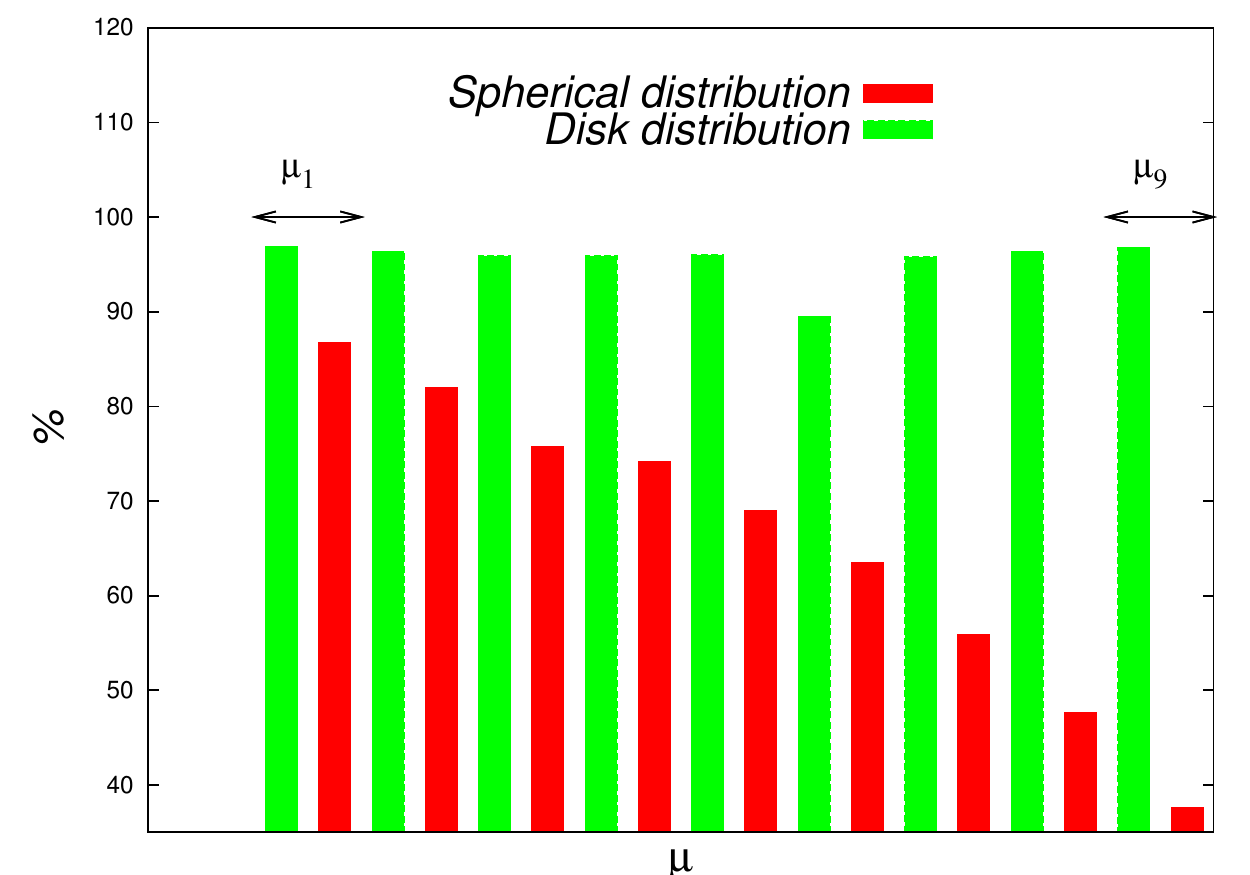}
\caption{Accuracy in the prediction set in terms of the inclination parameter $\mu$ for the disk  case. For this plot we use the images with resolution of $32\times 32$ pixels.}
\label{fig:Dependence_spin}
\end{figure}

In Figure \ref{fig:Dependence_spin} we show the results of classification of the test set when using the disk and spherical matter distributions as a function of the orientation parameter $\mu$, for the image resolution $32\times 32$ pixels. This is the quantitative accuracy measure of the results in Figures \ref{fig:Dependence_mu} and \ref{fig:Dependence_spin_sph}. The SVM is very accurate for disks, except for orientation in the subset $\mu_6$ where the accuracy of of 89.47\%. The accuracy in prediction for the spherical distribution is not as good and degrades when increasing $\mu$. The accuracy drops below 40\% for  nearly top head orientations within the range $\mu_9$.

In order to grasp the order of accuracy of a good classification, according to the definition of the spin classes above, when the SVM classifies correctly, the uncertainty in the estimation of the spin parameter, is the size of the interval defining the classes, namely 0.198 which represents a 10\% error among the whole range of possible values of spin. For instance, if the SVM predicts that certain image is classified in class $C_1$, this means that the spin parameter associated with the black hole seen in that image is $-0.891\pm0.099$. The same applies for the other nine classes.

{\it Classification of matter distribution.} In a real case scenario, it is expected that the image of a black hole does not exactly correspond to a theoretical matter distribution model, and a method like a SVM will be trained with approximate models expected to predict a property of the black hole. In terms of the matter models used here, one can translate the question and ask about the accuracy in the classification of an image due to a spherical distribution by a SVM trained with disk-like distributions or vice-versa. In our case we have already shown that for disk distributions the SVM is very accurate, whereas for spherical distributions it shows considerable confusion at classifying, more importantly for higher values of $\mu$. Therefore, in order to keep the accuracy percentage illustrated in Fig. \ref{fig:Dependence_spin} we add a strategic step consisting of a classification of matter distribution prior to the spin classification. For this, we trained a SVM able to distinguish between images generated with disk distributions from those generated with spherical distributions for different values of $\mu$. 
The result is that we are able to distinguish 100\% of images generated with disks from those generated with spherical distributions for the range $\mu_1$ and 99.94\% for the range $\mu_9$ calculated with the prediction sets. In a more general case, a wider variety of morphologies should be used to classify.
In the end, the final accuracy associated to the whole process of classification is the product of the morphology   and spin classification accuracies.

% -----> CONCLUSIONS
\section{Conclusions}
\label{sec:five}

In this paper we study the classification of the spin parameter of a black hole analyzing the images generated by two different distributions of matter around the black hole in a single wavelength. We show the performance of a SVM at classifying the spin as function of the image resolution and angle of vision. The results for disk distributions are very accurate, independently of the angle of vision, whereas for the spherical distribution the results are less accurate, and highly dependent on image resolution and angle of vision.

The accuracy presented in this manuscript is achieved using ten classes corresponding to uncertainties of 10\% on the estimation of the spin parameter. In order to reduce such uncertainty it would be necessary to increase the number of classes which can be done by increasing the number of generated images to ensure that the training process of the support vector machine is done properly.

The results presented here correspond to extremely poor image resolutions that could be much smaller than those  obtained by the EHT for Sgr A${}^{*}$, nevertheless they might well be suitable for SMBHs in other host galaxies. In a next step of our studies, we expect to include more realistic matter and radiation models based on numerical simulations of accretion flow processes and explore multi-band emissions.

% ----->     ACKNOWLEDGMENTS  

\section*{Acknowledgments}
This research is supported by Grants CIC-UMSNH-4.9, CIC-UMSNH-4.23, CONACyT 258726 
(Fondo Sectorial de Investigaci\'on para la Educaci\'on). The runs  were carried out in the computer farm funded by CONACyT 106466 and the Big Mamma cluster at the IFM.

% ----->     REFERENCES     

\end{document}